\documentclass[osajnl,twocolumn,showpacs,superscriptaddress,10pt]{revtex4-1}
\usepackage{amsmath,amssymb,graphicx}
\usepackage[center]{subfigure}

\begin{document}

\title{Buffering and Trapping Ultrashort Optical Pulses in Concatenated
Bragg Gratings}
\author{Shenhe Fu$^{1}$, Yikun Liu,$^{\dag,1}$ Yongyao Li$^{2,3}$, Liyan
Song,$^{1}$ Juntao Li,$^{1}$ Boris A. Malomed,$^{3}$ and Jianying Zhou}
\address{State Key Laboratory of Optoelectronic Materials and Technologies,
Sun Yat-sen University, Guangzhou 510275, China\\
$^{2}$Department of Applied Physics, South China Agriculture
University, Guangzhou 510642, China\\
$^{3}$Department of Physical Electronics, School of Electrical
Engineering,Faculty of Engineering, Tel Aviv University, Tel Aviv
69978, Israel \\
$^{\dag}$Corresponding author: ykliu714@gmail.com}
\begin{abstract}
Strong retardation of ultrashort optical pulses, including their
deceleration and stoppage in the form of Bragg solitons in a cascaded
Bragg-grating (BG) structure, is proposed. The manipulations of the pulses
are carried out, using nonlinear effects, in a chirped BG segment which is
linked, via a defect, to a uniform grating. The storage of the ultrashort
pulses is shown to be very robust with respect to variations of the input
field intensity, suggesting the feasibility of storing ultrafast optical
pulses in such a structure. Physical estimates are produced for the BGs
written in silicon.\newline
\end{abstract}

\maketitle

\begin{figure}[tbh]
\centering
\subfigure[]{\includegraphics[width=4.2cm, height=3.2cm]{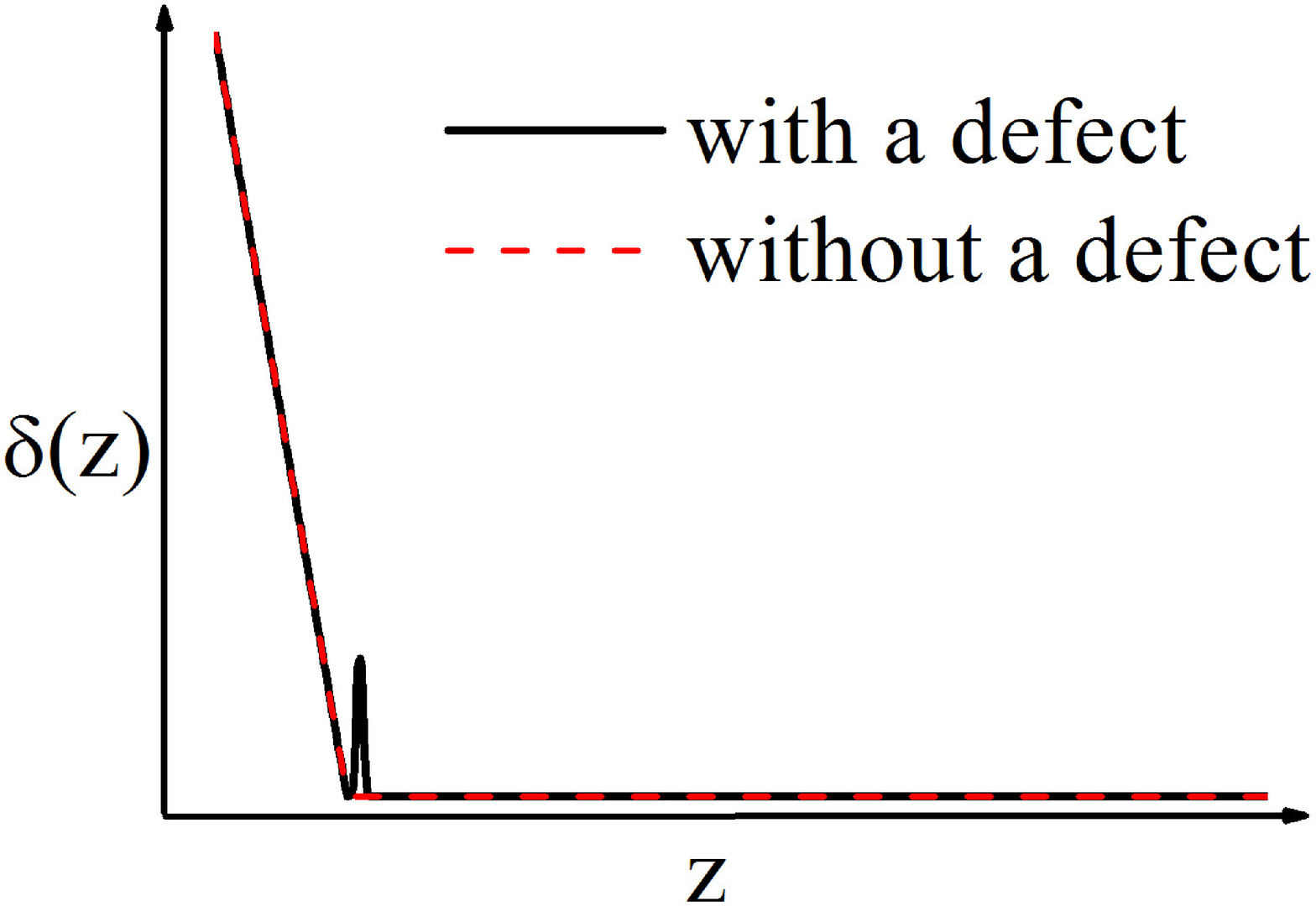}} %
\subfigure[]{\includegraphics[width=4.2cm, height=3.2cm]{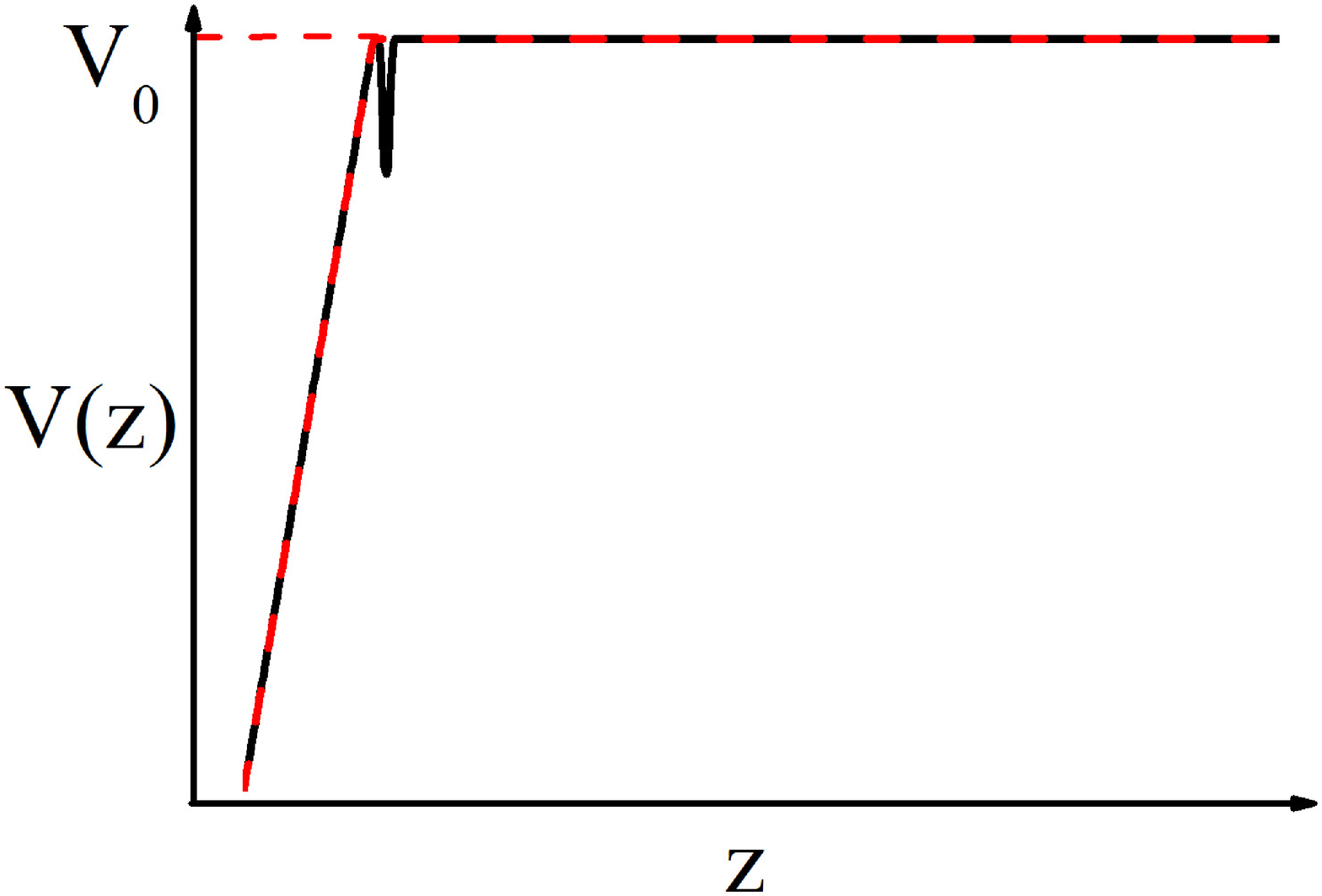}} %
\subfigure[]{\includegraphics[width=4.2cm, height=3.2cm]{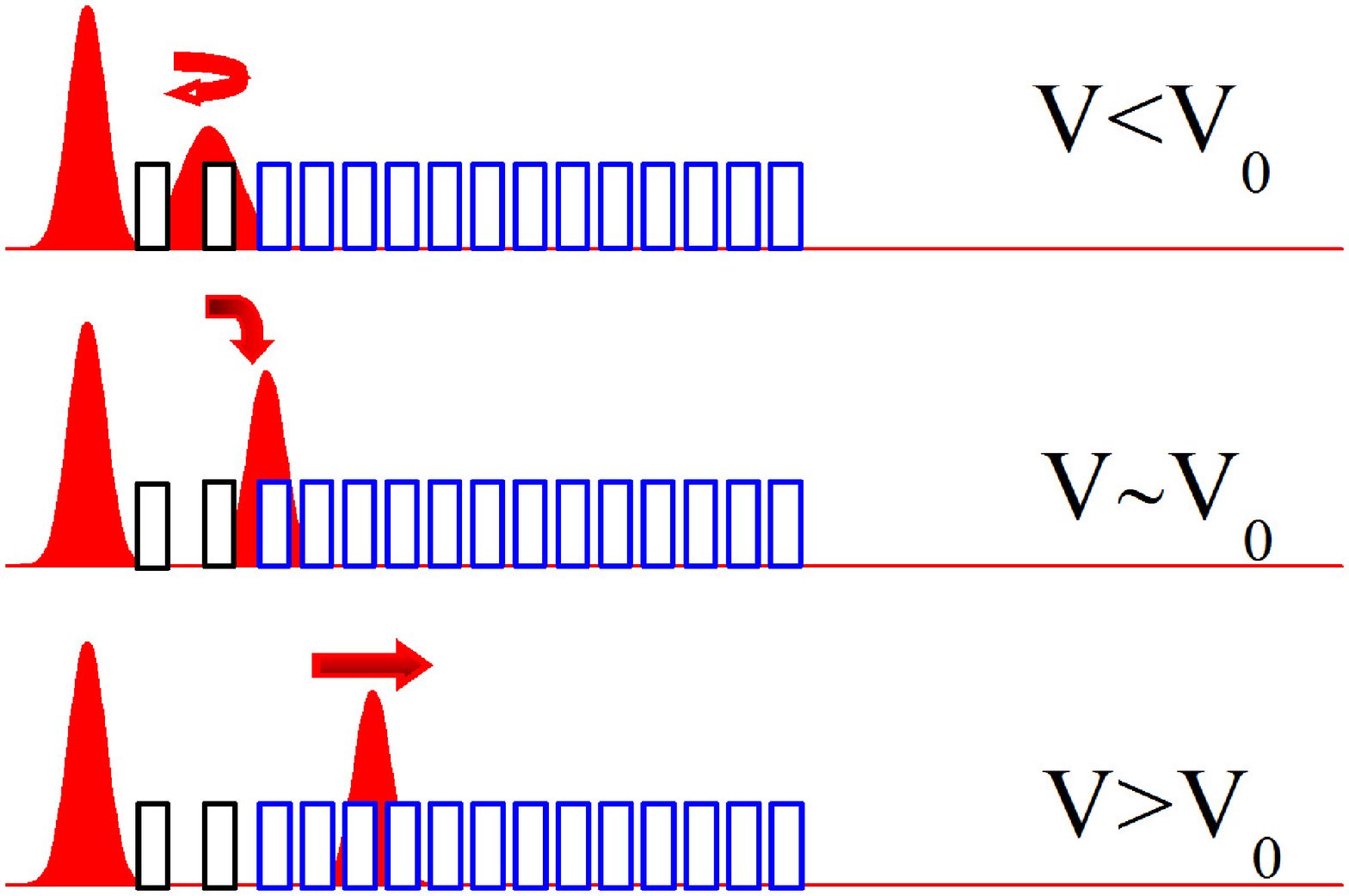}}
\subfigure[]{\includegraphics[width=4.2cm, height=3.2cm]{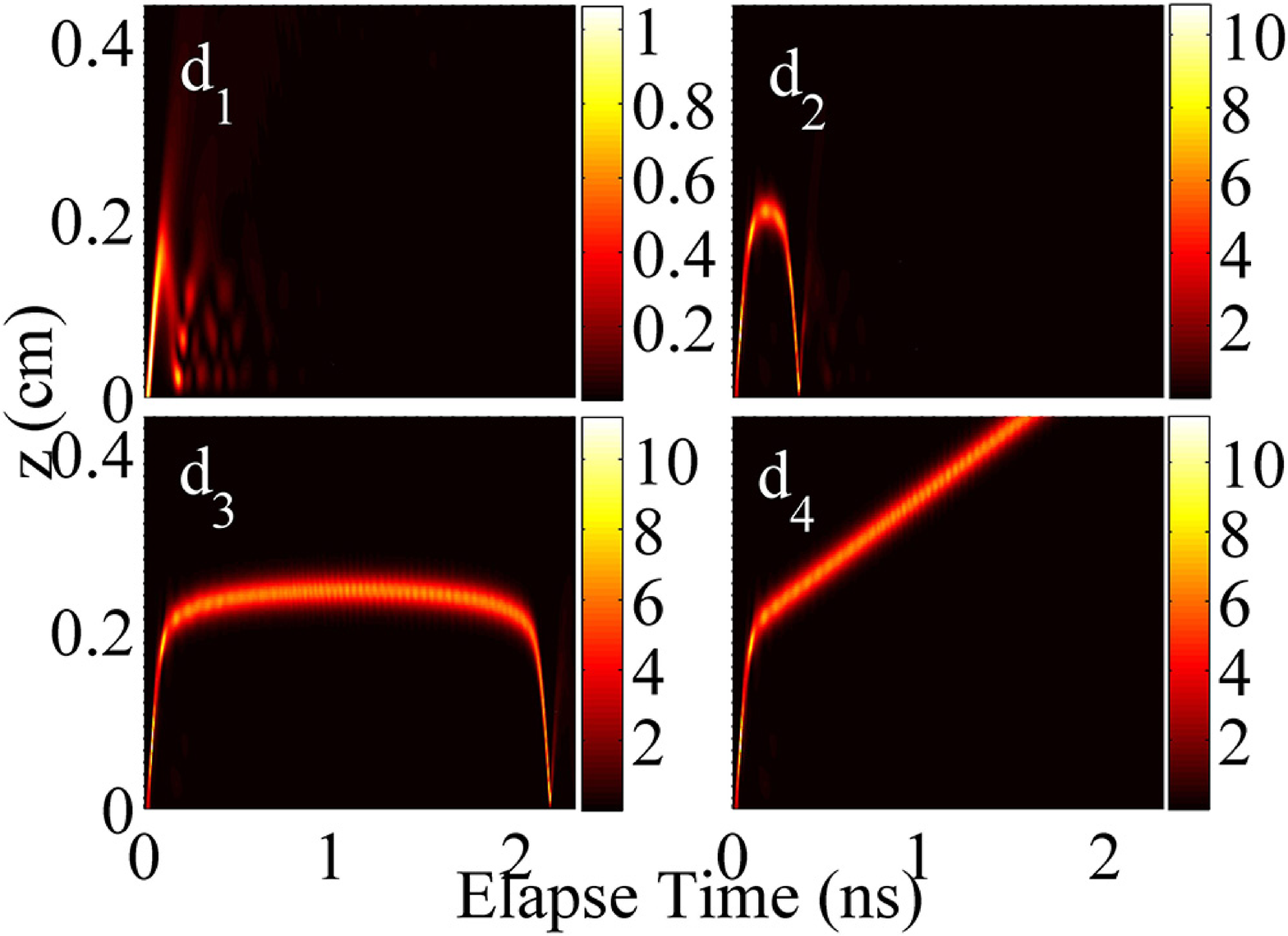}}
\caption{(Color online) (a) The wavenumber detuning as a function of the
propagation distance. (b) The dependence of the effective potential energy
on the propagation distance, which is reverse to that of the detuning. (c) A
sketch of the pulse propagation in the gratings without the defect, in three
different cases. (d) Simulations of the pulse propagation for different peak
intensities, injected into the grating without the defect: ($d_{1}$) 0.65
GW/cm$^{2}$, ($d_{2}$) 2.26 GW/cm$^{2}$, ($d_{3}$) 2.29 GW/cm$^{2}$, ($d_{4}$%
) 2.30 GW/cm$^{2}$.}
\label{fig1}
\end{figure}

\noindent Using the slow or stopped light to buffer optical signals has
drawn a great deal of attention \cite{CJCH,JTMC,RLME} due to potential
applications to optical telecommunication and signal processing, as well as
implications for studies of fundamental properties of optical pulses
propagating in complex media. Various schemes have been proposed for making
slow light, including electromagnetically induced transparency \cite%
{AKMJ,CLZD,JBK}, photonic-crystal waveguides \cite{YAVM,HAJV,TB,JLLS},
resonantly-absorbing gratings \cite{CJCH,AEKG,JTLJ,WNXJ} and resonant fiber
loops \cite{RLME}, among others \cite{RWB,Khurgin2}. However, for high-speed
signal processing and communications operating with short input pulses at
high repetition rates \cite{BCCM,CKPV}, the design of optical-signal
buffering is still a challenging problem.\newline
\indent In this context, designing devices for the generation of slow light
on the basis of Bragg-grating (BG) structures \cite%
{JTMC,JABA,DRNJ,DNCR,BJER,PYPC} is of great interest due to their tunability
and cascadability. In such periodic structures, strong dispersion of light
is induced in the vicinity of the photonic bandgap \cite{JTMC}. Mobile and
immobile BG solitons were predicted theoretically long ago \cite{DNCR,ABAS}.
A possibility of their retardation using apodized gratings \cite{WCKM}, as
well as producing standing soliton-like pulses as a result of the collisions
of counterpropagating moving ones \cite{WCKMB}, were analyzed in detail too.
Experimentally, BG solitons were first created with the velocity at 75\% of
the speed of light in the uniform optical fiber in the cladding of which the
BG was written \cite{BJER}. Later, slow robust light pulses traveling at
16\% \cite{JTMC} and 19\% \cite{KQLZ} of this speed were reported.
Theoretical and experimental studies of slow light in chirped BGs \cite{ENTC}
and superstructures \cite{DJGG} have also been performed. Nevertheless,
standing or very slow light in BGs has not been reported yet, which remains
a fundamental issue.\newline
\indent Generally, two conditions should be met for the creation of very
slow BG solitons. Firstly, the grating-induced dispersion must be balanced
by the Kerr nonlinearity \cite{CMSJ}. Secondly, since the optical field is
represented by forward- and backward-traveling waves in the BG, an initial
configuration should be built with the nearly equal powers of the two waves
\cite{CMS}. The former condition can be achieved using a high-power input,
whose nonlinearity is strong enough \cite{BJER}. The latter condition is
much harder to meet with the single incident pulse \cite{RSRS}, as it does
not initially contain any backward component.\newline
\indent The objective of this work is to explore possibilities for the
generation of slow ultrashort pulses, or even standing ones, in concatenated
BGs, built as a linearly chirped grating which is linked, through a defect,
to a uniform one. Essential advantages of this setting are demonstrated
below. Firstly, initial conditions for pulses at the input edge of the
uniform-BG segment may be manipulated by means of the preceding chirped BG,
which provides a possibility of preparing the right mix of the forward and
backward fields. Secondly, stable standing light pulses trapped by the
defect at the junction between the chirped and uniform segments are found in
a broad interval of values of the light intensities. Without the defect, the
stopped pulses turn out to be unstable.\newline
\indent We assume that the period of the chirped BG decreases continuously
along the propagation distance, while the period of the uniform BG is fixed
to be equal to its value at the output edge of the chirped BG. Accordingly,
the nominal Bragg wavelength is gradually shifted to smaller values for the
pulse propagating along the chirped BG, hence the wavelength of the input
pulse, originally taken near the blue edge of the bandgap, will be slightly
shifted into the depth of the bandgap. As a result, a fraction of the power
of the forward field component is converted into the backward field, leading
to a decrease of the pulse's velocity. Selecting an appropriate chirp, and
making use of the Kerr nonlinearity, it is possible to create conditions for
the generation of very slow or even immobile pulses.\newline
\indent First, we consider the setting without the grating defect, hence the
linear refractive index of such a tailored BG takes the form of \cite{LP} $%
n(z)=n_{0}\left[ 1+2\Delta n\cos \left( 2\pi z\left( 1+Cz\right) /\Lambda
_{0}\right) \right] $ for $0\leq z<L/8$, and $n(z)=n_{0}\left[ 1+2\Delta
n\cos \left( 2\pi z\left( 1+(1/8)CL\right) /\Lambda _{0}\right) \right] $
for $L/8\leq z\leq L$, where $z$ is the propagation distance, $L$ the total
length of the grating, $n_{0}$ the average refractive index, $\Delta n$ its
modulation depth, $\Lambda _{0}$ the BG period at the input edge, and $C$ is
the chirp. The implication of this setting is that the length of the chirped
segment is $1/8$ of the total length. As said above, the chirped segment is
introduced for manipulating the form of the pulse arriving at the input edge
of the uniform grating.\newline
\indent The standard coupled-mode theory \cite{CMSJ,LP} for the light
propagation in uniform BGs can be applied as well to chirped gratings \cite%
{ENTC} and Bragg superstructures \cite{DJGG}. For the slowly varying
envelopes of forward and backward waves, $E_{f}$ and $E_{b}$, the
coupled-mode equations are written as \cite{LP}
\begin{gather}
\pm i\frac{\partial E_{f,b}}{\partial z}+\frac{i}{v_{g}}\frac{\partial
E_{f,b}}{\partial t}+\delta (z)E_{f,b}+\kappa (z)E_{b,f}  \notag \\
+\gamma (|E_{f,b}|^{2}+2|E_{b,f}|^{2})E_{f,b}=0,  \label{couple-mode}
\end{gather}%
where $t$ is time, $v_{g}=c/n_{0}$ is the group velocity in the material of
which the BG is fabricated, $c$ is the speed of light in vacuum, $\gamma
=n_{2}\omega /c$ is the nonlinearity strength, with $\omega $ being the
frequency of the carrier wave and $n_{2}$ the Kerr coefficient. The
wavenumber detuning parameter in Eqs. (\ref{couple-mode}), corresponding to
the adopted BG profile, is $\delta (z)=\Delta -2\pi Cz/\Lambda _{0}$ for $%
0\leq z<L/8$, and $\delta (z)=\Delta -2\pi CL/\left( 8\Lambda_{0} \right)$
for $L/8\leq z\leq L$, where $\Delta \equiv 2\pi n_{0}/\lambda -\pi /\Lambda
_{0}$, while $\kappa =\pi \Delta n/\Lambda _{0}$ represents the coupling
between the forward and backward fields.\newline
\indent The variation of the wavenumber-detuning $\delta (z)$, as a function
of the propagation distance, $z$, is shown in Fig. \ref{fig1}(a). If Eqs. (%
\ref{couple-mode}) are considered as coupled Schr\"{o}dinger equations, the
corresponding effective potential energy $V(z)$ is proportional to $-$ $%
\delta (z)$ [as shown in Fig. \ref{fig1}(b)], which may be used for the
qualitative analysis of the motion of the pulse treated as a quasiparticle
\cite{WCKMBA}. Accordingly, if the initial kinetic energy of the
quasiparticle, $V$, is smaller than the asymptotic level $V_{0}$, the moving
pulse bounces back, as illustrated in the top plot of Fig. \ref{fig1}(c).
With the initial energy $V=V_{0}$ [the middle plot in Fig. \ref{fig1}(c],
the pulse nearly comes to a halt, which is explained by the character of the
coupling between the forward and backward fields in the chirped BG, and its
interplay with the Kerr nonlinearity. Finally, a pulse slowly advancing into
the uniform BG can be obtained for the initial energy slightly exceeding $%
V_{0}$ [the bottom plot in Fig. \ref{fig1}(c)]. Below, these three options
are confirmed by systematic simulation of Eqs. (\ref{couple-mode}).

The simulations were performed with physical parameters of silicon, whose
self-focusing-Kerr coefficient is $4.5\times 10^{-14}$ cm$^{2}$/W \cite{MDFQ}%
[here, we neglect the two-photon absorption (TPA) effect, which is essential
in silicon \cite{pulsed}, and is included below], and the average refractive
index is $n_{0}=3.42$. The refractive-index modulation depth and chirp are
taken as $\Delta n=0.006$ and $C=2.5064\times 10^{-4}$ cm$^{-1}$,
respectively. The sufficient total length $L$ is $1.5$ cm. At the input edge
of the sample, the BG period is fixed to be $\Lambda _{0}=154.1$ nm, and the
Gaussian pulse, of temporal width $16$ ps, is coupled into the system at
carrier wavelength $1053$ nm.

\begin{figure}[t!]
\centering
\subfigure[]{\includegraphics[width=4.2cm, height=3.2cm]{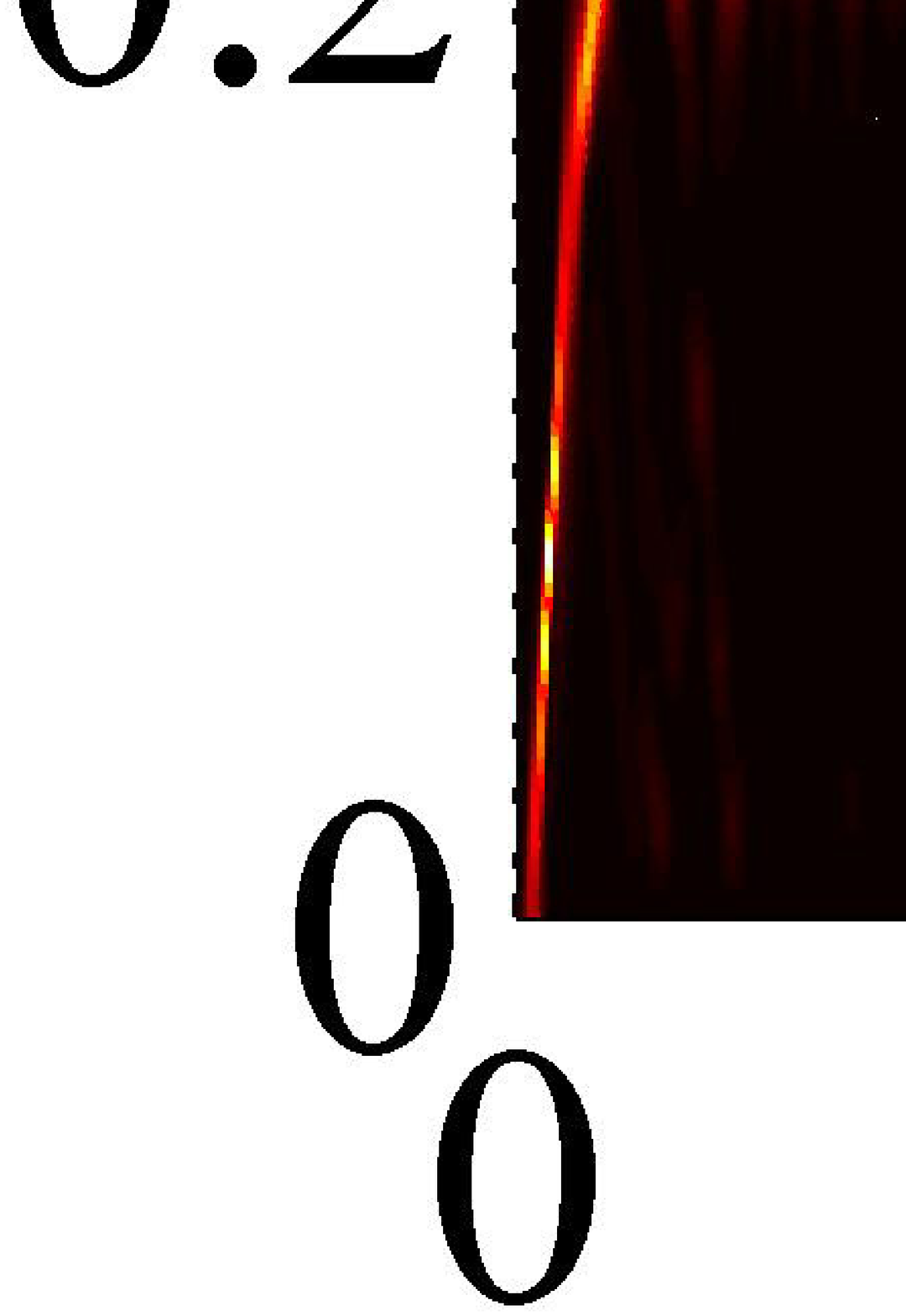}}
\subfigure[]{\includegraphics[width=4.2cm, height=3.2cm]{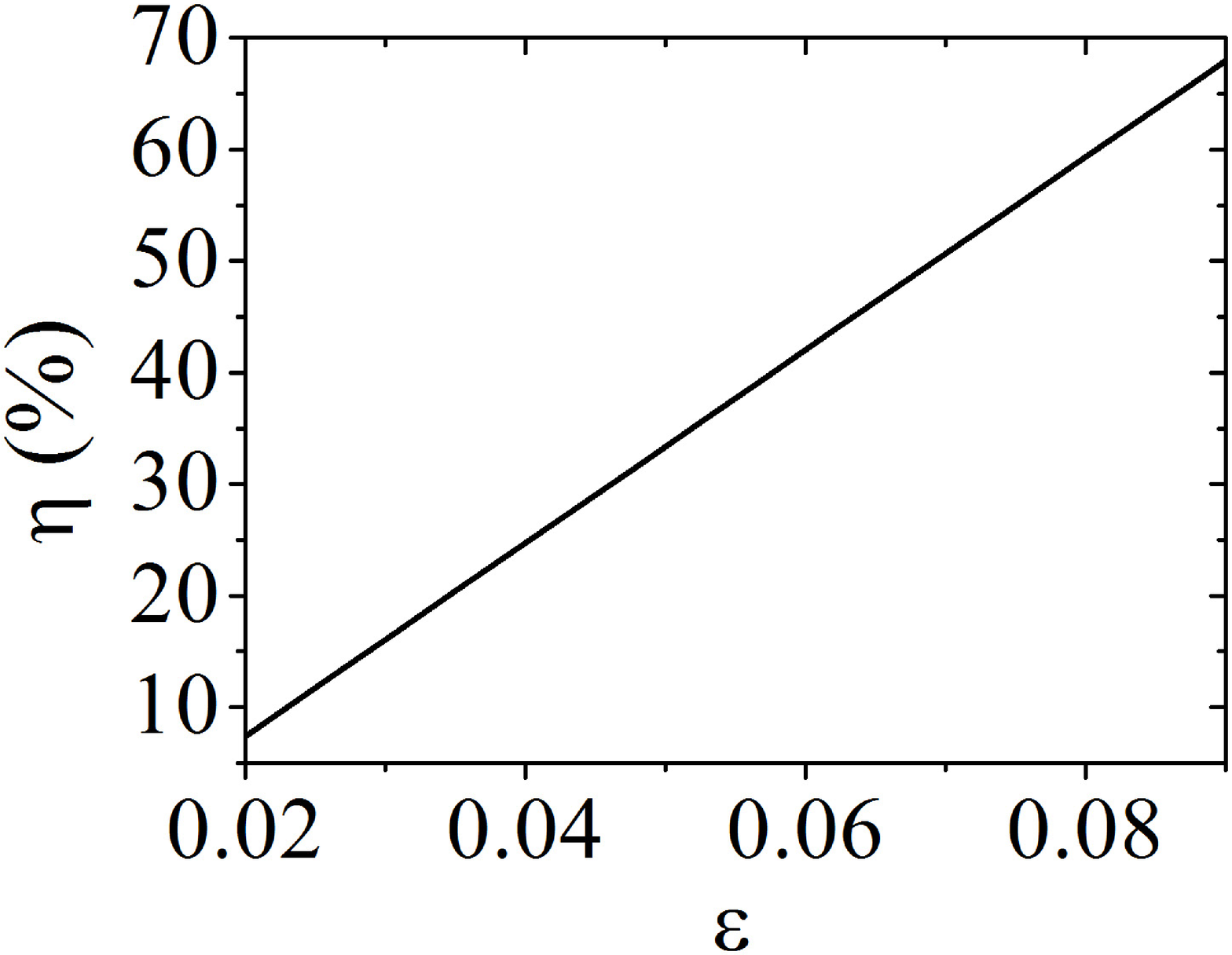}}
\caption{(Color online) (a) Examples of trapping the incident pulse by the
defect with strength $\protect\varepsilon =0.05$, at the following values of
the initial peak intensity, $I_{P}$: $1.94$ GW/cm$^{2}$ (a$_{1}$) , $2.14$
GW/cm$^{2}$ (a$_{2}$) , $2.33$ GW/cm$^{2}$ (a$_{3}$), $2.59$ GW/cm$^{2}$ (a$%
_{4}$). (b) The relative width of the trapping interval as a function of the
defect's strength, $\protect\varepsilon $.}
\label{fig2}
\end{figure}

With these parameters, simulations of Eq. (\ref{couple-mode}) produce the
results displayed in Fig. \ref{fig1}(d). From these results, three
conclusions can be made, varying the tunable input-pulse peak intensity, $%
I_{P}$ -- first, in the absence of the inserted defect.\newline
\indent(i) At low intensities, such as $I_{P}=0.65$ GW/cm$^{2}$ [panel (d$%
_{1}$) in Fig. \ref{fig1}(d)], the pulse is almost totally reflected by the
photonic bandgap, conspicuously broadening in the course of the propagation
through the chirped BG, i.e., the pulse's dispersion is not compensated by
the weak nonlinearity.\newline
\indent(ii) An unstable standing light pulse trapped at the interface is
generated at higher intensities. In particular, at $I_{P}=2.26$ GW/cm$^{2}$,
the pulse stops for a while and is then reflected, as seen in panel d$_{2}$
of Fig. \ref{fig1}. Slightly increasing $I_{P}$ to 2.29 GW/cm$^{2}$ in panel
d$_{3}$, it is observed that the pulse halts at the interface for almost 1.3
ns, but eventually shows its intention to start the reverse motion. The
pulse's peak intensities in this case, ranging from 2.26 GW/cm$^{2}$ to 2.29
GW/cm$^{2}$, slightly change in the course of the evolution.\newline
\indent(iii) A very slow pulse moving into the uniform BG, which may be
identified as a BG soliton, can be created, as shown in panel d$_{4}$, for $%
I_{P}=2.30$ GW/cm$^{2}$. In this case, the pulse is propagating in the
uniform BG with a very small velocity, $0.005c$ (which is much smaller than
the results obtained in Refs. \cite{JTMC,BJER,KQLZ}). Naturally, the
simulations demonstrate that the velocity of the moving soliton increases
with the further increase of $I_{P}$.\newline
\indent Since defects may help in trapping light in the Bragg gratings \cite%
{PYPC,RHGR,IVKI}, we now modify the setting, introducing a local defect
between the chirped and uniform BG segments, as shown in Fig.\ref{fig1}(a).
The defect is defined by taking the wavenumber detuning as $\delta
^{^{\prime }}(z)=\delta (z)\left\{ 1+\varepsilon ~\mathrm{sech}\left[
-\left( \left( z-L/8\right) /0.005\right) ^{2}\right] \right\} $, where $%
\varepsilon $ is the strength of the defect, and its width is taken as $50$ $%
\mathrm{\mu }$m. In comparison with the spatial width of the pulse (the
temporal duration of $16$ ps corresponds to $4.8$ mm in space) or the total
length, $L $, the width of 50 $\mathrm{\mu }$m may indeed be considered as
that of a point-like defect. The defect affects the shape of the effective
potential, adding a local trapping well to it, as shown in Fig. \ref{fig1}%
(b).\newline
\indent Noteworthy effects produced by simulations of the setting including
the defect are presented in Fig. \ref{fig2}(a), for the defect's strength $%
\varepsilon =0.05$. The pulse may be \emph{stably} trapped by the defect (in
particular, at peak intensities $I_{P}=$ $2.33$ GW/cm$^{2}$ and $2.59$ GW/cm$%
^{2}$, at which the BG soliton would be moving into the uniform BG in the
absence of the defect). The stable trapping is observed in the interval of
the peak intensity with relative width $\eta =(I_{\mathrm{max}}-I_{\mathrm{%
min}})/(I_{\mathrm{max}}+I_{\mathrm{min}})/2\approx 33.2\%$, where $I_{%
\mathrm{max}}$ and $I_{\mathrm{min}}$ are the maximum and minimum value of $%
I_{P}$ for the generation of standing pulse, respectively.\newline
\indent The influence of the defect's strength $\varepsilon $ on the
trapping-interval's width $\eta $ was investigated too. As seen in Fig. \ref%
{fig2}(b), $\eta $ grows linearly with $\varepsilon $. This result is quite
natural, as the effective energy of the pulse approaching the defect-induced
trapping potential well is proportional to $I_{P}$, and, on the other hand,
the depth of the of potential well is proportional to $\varepsilon $. The
defect with $\varepsilon <0.01$ cannot trap the pulse, i.e., $\eta
(\varepsilon \leq 0.01)=0$. Additional simulations demonstrate that, for the
given setting, the decrease of the pulse's temporal width leads to shrinkage
and eventually vanishing of the stable-trapping interval, which is a natural
consequence of the fact that the spectral width of the pulse gets too broad
in comparison with the BG's bandgap.
\begin{figure}[t!]
\centering
\subfigure[]{\includegraphics[width=4.2cm,height=3.2cm]{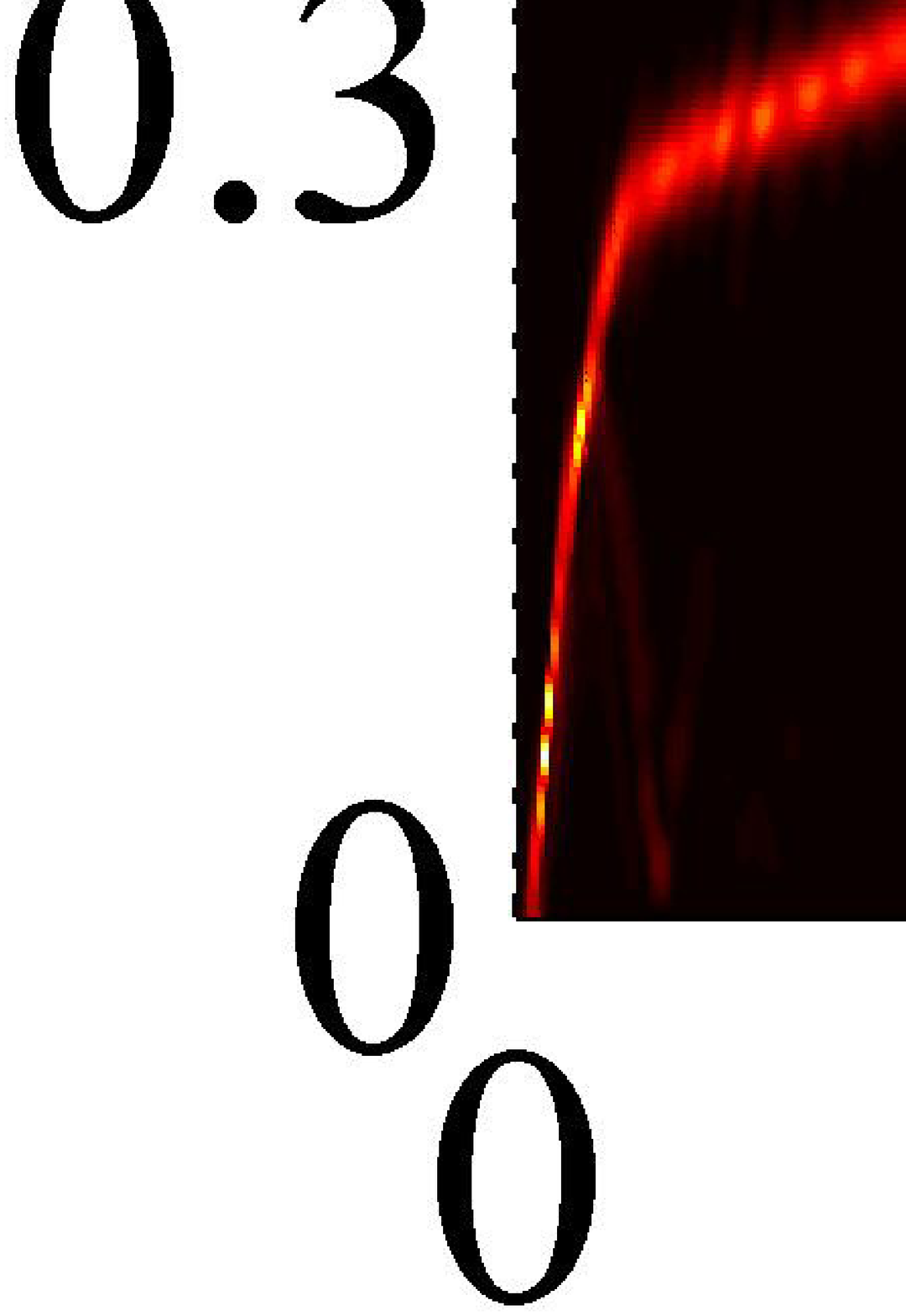}} %
\subfigure[]{\includegraphics[width=4.2cm,height=3.2cm]{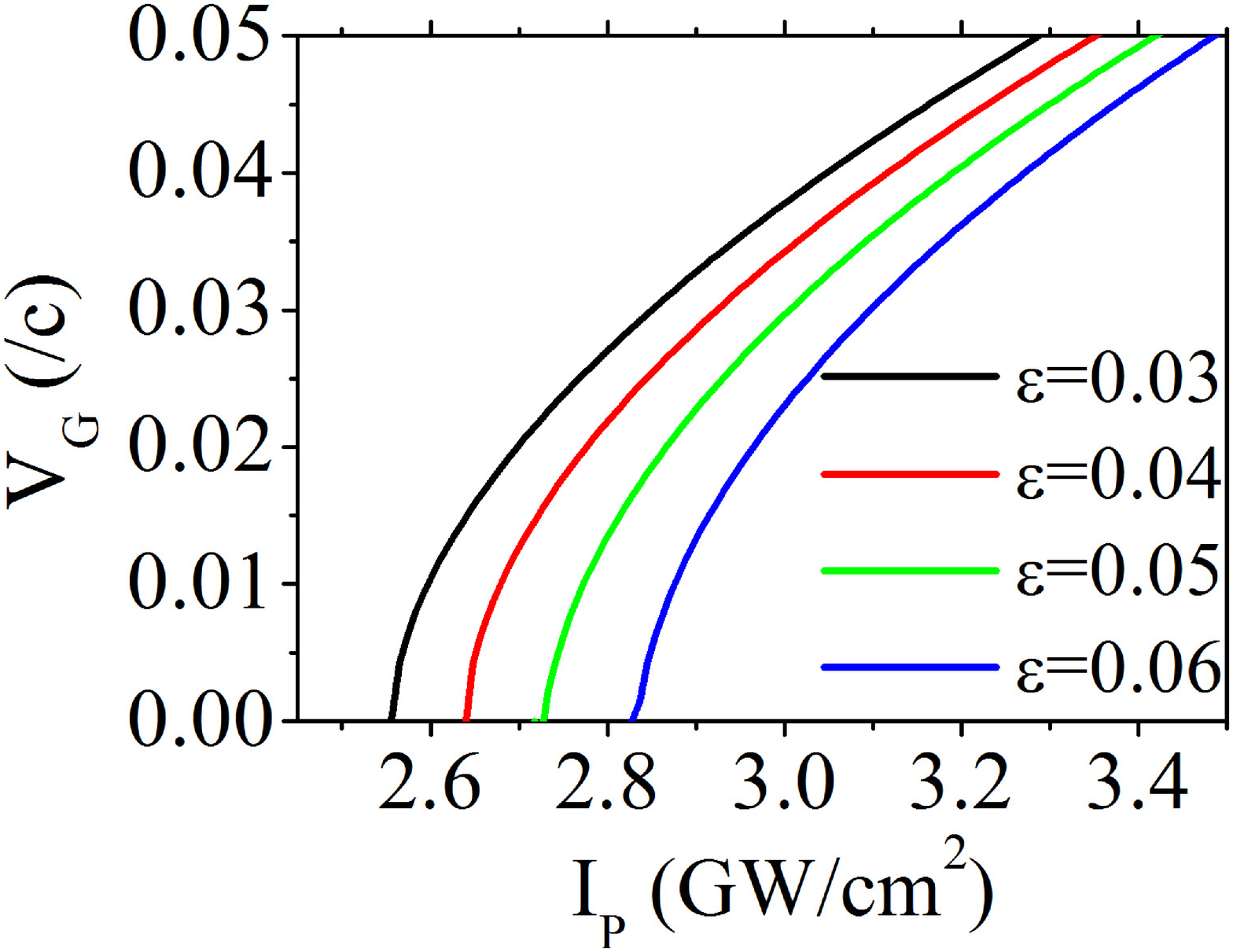}}
\caption{(Color online) Free motion of the Bragg soliton in the uniform
grating. (a) Trajectories of the motion: (a$_{1}$) for $\protect\varepsilon %
=0.04$, $I_{P}=2.65$ GW/cm$^{2}$; (a$_{2}$) for $\protect\varepsilon =0.04$,
$I_{P}=2.69$ GW/cm$^{2}$; (a$_{3}$) for $\protect\varepsilon =0.05$, $%
I_{P}=2.74$ GW/cm$^{2}$; (a$_{4}$) for $\protect\varepsilon =0.05$, $%
I_{P}=2.78$ GW/cm$^{2}$. (b) The soliton's velocity versus the initial peak
intensity, $I_{P}$, at different values of the defect's strength.}
\label{fig3}
\end{figure}
\begin{figure}[t!]
\subfigure[]{\includegraphics[width=4.2cm, height=3.2cm]{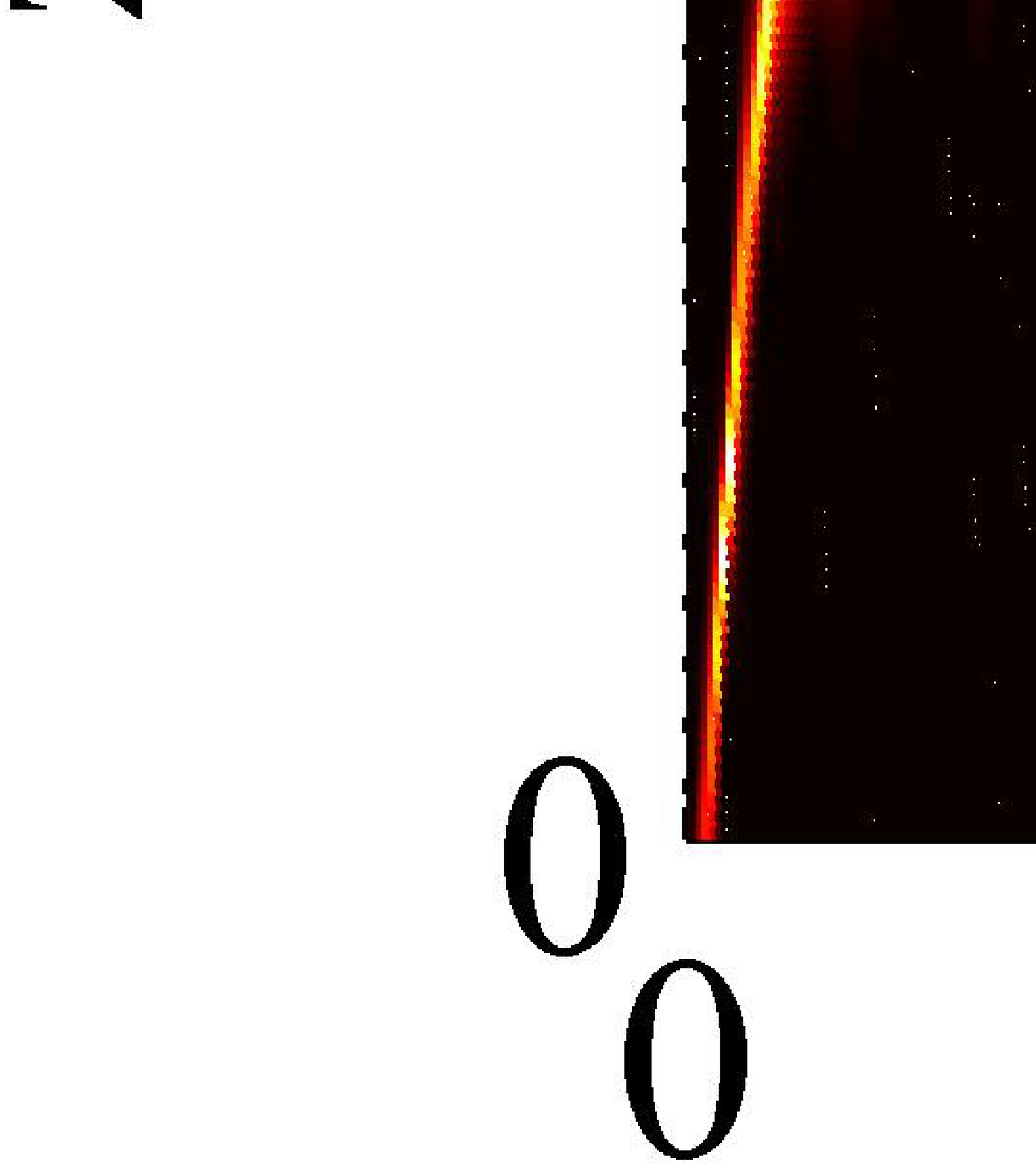}} %
\subfigure[]{\includegraphics[width=4.2cm, height=3.2cm]{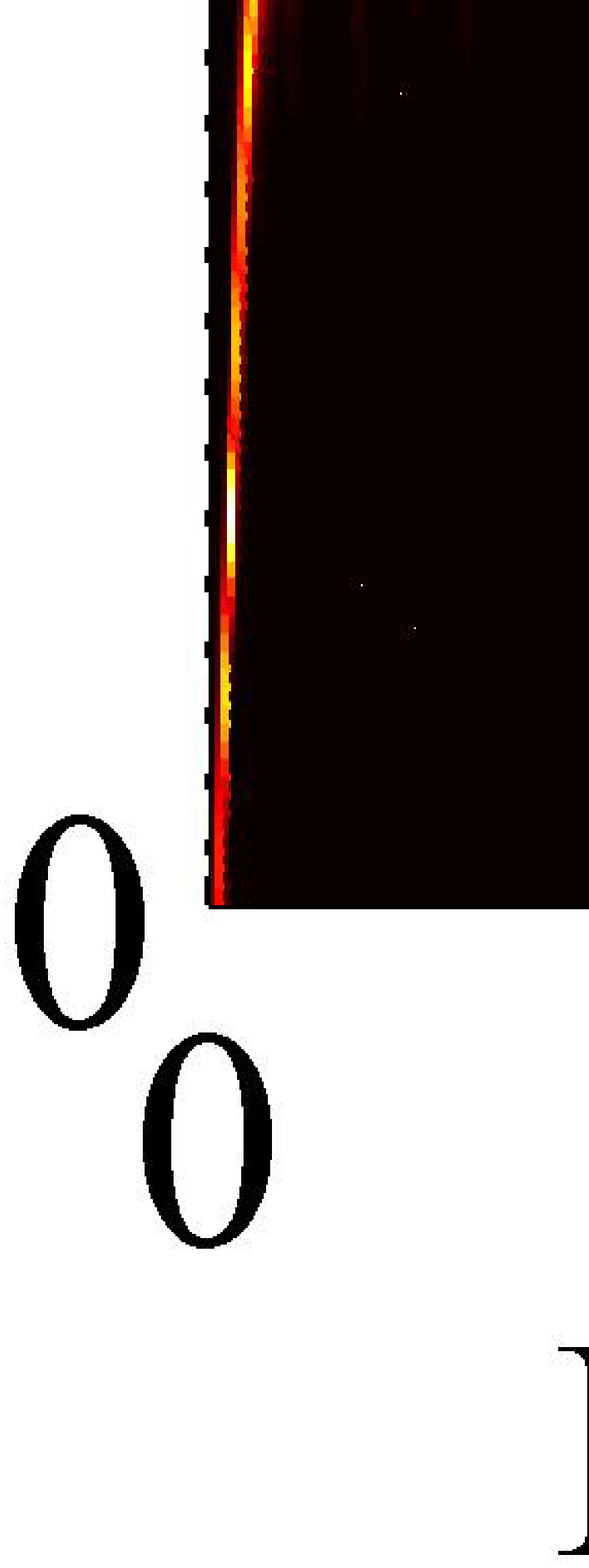}} %
\caption{(Color online) A typical example of the stable trapping optical
pulse with $I_{P}=6.47$GW/cm$^{2}$, at $\protect\lambda =2$ $\mathrm{\protect%
\mu }$m, $C=2.6213\times 10^{-4}$ cm$^{-1}$, $\Lambda _{0}=292.7$ nm: (a)
without the TPA effect; (b) with the TPA effect included (Im[$n_{2}$]=0.2$%
\times$ 10$^{-16}$ cm$^2$/W).}
\label{fig4}
\end{figure} \newline
\indent To observe the Bragg soliton-like pulses passing the attractive defect and
moving into the uniform BG, a higher initial peak power is needed. For
instance, in Fig. \ref{fig3}(a), with defect's strength $\varepsilon =0.04$,
and peak intensities $I_{P}=$ $2.65$ GW/cm$^{2}$ and $2.69$ GW/cm$^{2}$
(panels a$_{1}$ and a$_{2}$), the pulses keep moving slowly with velocities
at $0.005c$ and $0.011c$, respectively. If $\varepsilon $ increases to $0.05$%
, similar results are obtained for $I_{P}=2.74$ GW/cm$^{2}$ and $2.78$ GW/cm$%
^{2}$, as seen in panels a$_{3}$ and a$_{4}$.\newline
\indent Generally, the slow motion of the Bragg soliton in the uniform BG
occurs, with or without the defect, when $I_{P}$ exceeds a certain threshold
value, $I_{P}^{\mathrm{(thr)}}$. To further elucidate the influence of the
defect on the pulse's dynamics, the dependence of velocity $V_{G}$ of the
soliton moving inside the uniform BG and $I_{P}$ is depicted in Fig. \ref%
{fig3}(b), for different values of $\varepsilon $. It is evident that,
slightly above the threshold, i.e., at small values of $I_{P}-I_{P}^{\mathrm{%
(thr)}}$, the velocity grows as $\sqrt{I_{P}-I_{P}^{\mathrm{(thr)}}}$, which
is typical for depinning phenomena (as the kinetic energy is proportional to
the squared velocity).\newline
\indent While, as mentioned above, silicon demonstrates the TPA in the
considered spectral region, these losses quickly decrease at $\lambda >$ $%
1700$ nm \cite{XLRM}. A typical example of stable trapping of the
soliton-like pulse with the carrier wavelength $\lambda =2$ $\mathrm{\mu }$%
m, predicted by simulations of Eq. (\ref{couple-mode}) with accordingly
modified parameters, is shown in Fig. 4(a). In particular, with the TPA
effect included (in this situation, we use a relevant value of its strength,
$\mathrm{Im}\left[ n_{2}\right] =0.2\times 10^{-16}$ cm$^{2}$/W), Fig. 4(b)
shows that the optical pulse trapped by the defect is maintained well.
Further simulations demonstrate that the storage time $\tau $ is nearly inversely proportional to the TPA
coefficient, measured in units of $10^{-16}$ cm$^{2}$/W: $\tau \left[
\mathrm{ns}\right] \approx 0.8\left( \mathrm{Im}\left[ n_{2}\right] \right)
^{-1}$. If necessary, $\tau $ can be made indefinitely long by adding a weak
local gain compensating the residual loss \cite{Mak}. \newline
\indent In conclusion,
we have performed the analysis of the generation of very slow or even
standing ultrashort optical pulses in concatenated BGs (Bragg gratings),
built as a linearly chirped grating linked to a uniform one, with a defect
placed at the junction between them. Physical parameters were taken for
silicon, taking care to decrease the detrimental effect of the TPA
(two-photon absorption). Systematic simulations have demonstrated that, in
agreement with our qualitative analysis, it is possible to provide suitable
conditions for the creation of very slow Bragg solitons in this system, or
even solitons stably trapped by the defect. The chirped BG segment is
necessary for the preparation of an appropriate pulse, built as the right
mix of forward and backward wave components, which impinges on the defect
and then advances into the uniform grating, unless it gets trapped by the
defect. The interval of input peak intensities of the light pulses, which
are eventually trapped by the defect, increases linearly with the defect's
strength. \newline
\indent This work is supported by Chinese National Basic Research Program (2010CB923204)
and Chinese National Natural Science Foundation
(11104083, 11204386).

\end{document}